\documentclass{aa} 

\usepackage{graphicx}
\usepackage{txfonts}
\usepackage{textcomp}

\usepackage{natbib}
\bibpunct{(}{)}{;}{a}{}{,} % to follow the A&A style

\catcode`\@=11
\def\gsim{\ifmmode{\mathrel{\mathpalette\@versim>}}
    \else{$\mathrel{\mathpalette\@versim>}$}\fi}
\def\lsim{\ifmmode{\mathrel{\mathpalette\@versim<}}
    \else{$\mathrel{\mathpalette\@versim<}$}\fi}
\def\@versim#1#2{\lower 2.9truept \vbox{\baselineskip 0pt \lineskip 
    0.5truept \ialign{$\m@th#1\hfil##\hfil$\crcr#2\crcr\sim\crcr}}}
\catcode`\@=12

\begin{document}

   \title{The GIRAFFE Inner Bulge Survey (GIBS). I. Survey Description and a kinematical
map of the Milky Way bulge.
   \thanks{Based on observations taken with ESO telescopes at the La Silla Paranal 
   Observatory under programme IDs 187.B-909 and 089.B-0830.}}

\author{
 M. Zoccali\inst{1,2} 
 \and
 O. A. Gonzalez\inst{3}
 \and
 S. Vasquez\inst{1,3}
 \and 
 V. Hill\inst{4}
 \and
 M. Rejkuba\inst{5,6}
 \and           
 E. Valenti\inst{5}
\and 
A. Renzini\inst{7}
\and
A. Rojas-Arriagada\inst{4}
 \and
I. Martinez-Valpuesta\inst{8,9}
 \and
 C. Babusiaux\inst{10}   
 \and 
 T. Brown\inst{11}         
 \and
 D. Minniti\inst{1,2,12}
 \and 
 A. McWilliam\inst{13}
}

\institute{
Instituto de Astrof\'{i}sica, Facultad de F\'{i}sica, Pontificia Universidad Cat\'olica 
de Chile, Av. Vicu\~na Mackenna 4860, Santiago, Chile.
\email{mzoccali@astro.puc.cl}
\and
The Milky Way Millennium Nucleus, Av. Vicu\~{n}a Mackenna 4860, 782-0436 Macul, 
Santiago, Chile
\and
European Southern Observatory, A. de Cordova 3107, Casilla 19001, Santiago 19, Chile
\and
Laboratoire Lagrange (UMR7293), Universit\' de Nice Sophia Antipolis, CNRS, 
Observatoire de la C\^{o}te d'Azur, BP 4229, 06304, Nice Cedex 4, France
\and 
European Southern Observatory, Karl-Schwarzschild-Strasse 2, 85748, Garching, Germany
\and
Excellence Cluster Universe, Boltzmannstr. 2, D-85748, Garching, Germany
\and
INAF - Osservatorio Astronomico di Padova, vicolo dell'Osservatorio 5, 35122, Padova, 
Italy
\and
Max-Planck-Institut f\"ur Extraterrestrische Physik, Giessenbachstrasse, 85748 Garching, Germany
\and
Instituto de Astrof\'\i sica de Canarias, Calle V\'\i a L\'actea s/n, 38205 La Laguna, 
Tenerife, Spain
\and
GEPI, Observatoire de Paris, CNRS UMR 8111, Universit\'e Paris Diderot, F-92125, Meudon,
Cedex, France
\and
Space Telescope Science Institute, 3700 San Martin Drive, Baltimore, MD 21218, USA
\and
Vatican Observatory, V00120 Vatican City State, Italy
\and
The Observatories of the Carnegie Institute of Washington, 813 Santa Barbara Street, 
Pasadena, CA 91101-1292, USA
}
   \date{Received; accepted}

% \abstract{}{}{}{}{} 
% 5 {} token are mandatory
 
  \abstract
  % context (optional)
{The Galactic  bulge is a massive,  old component of the  Milky Way. It
is known to host a bar, and it has recently been demonstrated to
have a pronounced boxy/peanut structure  in its outer region.  Several
independent  studies suggest  the presence  of more  than one  stellar
populations  in  the  bulge,  with different  origins  and  a  relative
fraction changing across the bulge area.}
  % aims 
{This is the first of a series  of papers presenting the results of the
Giraffe  Inner Bulge  Survey,  carried  out at  the  ESO-VLT with  the
multifibre spectrograph FLAMES. Spectra of $\sim$5000 red clump giants
in 24  bulge fields have  been obtained  at resolution R=6500,  in the
infrared Calcium  triplet wavelength region at  $\sim$8500 \AA.  They
are used  to derive radial  velocities and metallicities, based on new
calibration specifically  devised for this project.  Radial velocities
for another $\sim$1200  bulge red clump giants,  obtained from similar
archive data, have been added to the sample. Higher resolution spectra
have  been  obtained  for   $\sim$450 additional stars  at  latitude
$b=-3.5$, with the aim of investigating chemical abundance patterns 
variations with longitude, across the inner bulge. In total
we present here radial velocities for 6392 red clump stars.}
  % methods 
{We present here the target selection criteria, observing strategy and
the catalogue with radial velocity measurements for all the target stars.}
  % results 
{We derive a radial velocity, and  velocity dispersion map of the Milky
Way bulge, useful to be compared with similar maps of external bulges,
and to  infer the expected  velocities and  dispersion at any  line of
sight.   The   K-type  giants   kinematics  is  consistent   with  the
cylindrical rotation  pattern of M-giants  from the BRAVA  survey. Our
sample enables to extend this result to latitude $b=-2$, closer to the
Galactic  plane than  probed  by previous  surveys.  Finally, we  find
strong  evidence  for  a  velocity dispersion  peak  at  ($0,-1$)  and
($0,-2$), possibly indicative  of a high density peak  in the central
$\sim$250 pc of the bulge.}
  % conclusions (optional) 
{}


   \maketitle
%
%___________________________________________________________________
\section{Introduction}

The Giraffe  Inner Bulge  Survey (GIBS)  is a  survey of
$\sim$ 6400 red  clump (RC) stars in the Milky  Way bulge, carried out
with the GIRAFFE spectrograph of the FLAMES instrument at the ESO Very
Large Telescope  (VLT). The aim  of the GIBS  survey is to  derive the
metallicity  and radial  velocity distributions of bulge  stars across
different  fields, spread  over a  large  area of  the inner  Galactic
bulge, that is also part of the VISTA Variable in the V\'{i}a L\'actea (VVV) 
survey \citep{minniti+2010}.

The inner region ($\lsim  3$ kpc) of the Milky Way  galaxy is known to
host a  bar \citep[][and references  therein]{blitz+1991, stanek+1994,
  dwek+1995,   babusiaux+2005,   rattenbury+2007}.    Only   recently,
however,   a  split   RC  was   discovered  at   $l$=0  and   $|b|$>5,
\citep{nataf+2010,  mcwilliam+2010} and  detailed 3D  maps constructed
using RC  stars as distance indicators  revealed that the bulge  is in
fact  X-shaped  \citep{saito+2011,  wegg+2013}.    The  X-shape  is  a
pronounced  boxy/peanut  (B/P) structure  that,  in  its inner  region
($|b|$<5), becomes a bar, hereafter the  {\it main} bar. Closer yet to
the center ($|l|$<4, $|b|$<2) the  {\it main} bar changes its apparent
inclination with  respect to the line  of sight \citep{nishiyama+2005,
  gonzalez+2011_bar},  most likely  due to  the presence  of either  a
distinct,   smaller   bar,   or    a   more   axysimmetric   structure
\citep{gerhard+2012}.   The presence  of  a longer  bar, extending  to
longitudes   $|l|$>7    \citep{hammersley+2000,   cabrera-lavers+2007,
  cabrera-lavers+2008,     lopez-corredoira+2007,     churchwell+2009,
  amores+2013} has  been interpreted  by theoreticians as  most likely
being    an    extension    of    the   {\it    main}    bar    itself
\citep[e.g.,][]{inma+2011, athanassoula+2012}.

The axial  ratio of the  {\it main} bar  is close to  1:0.35:0.25, but
rather different values  of the inclination angle with  respect to the
line of sight are found in the literature, spanning the range 15 to 45
degrees,  depending on  the method  used to  trace it  \citep[e.g.,][]
{binney+1997,      dehnen+2000,     bissantz+2002,      benjamin+2005,
  babusiaux+2005, rattenbury+2007, robin+2012, cao+2013}.

The  ultimate  goal, when  studying  the  Galactic  bulge, is  to  set
constrains  on the  formation mechanism(s)  of the  Milky Way  and, by
extension,  of  galaxies  in  general. Dynamical  models  predict  the
formation of B/P structures as the outcome of the secular evolution of
a disk, through the formation and
successive vertical  heating of a  bar \citep[e.g.,][]{pfenniger+1991,
  athanassoula+2005,  inma+2006, debattista+2006}.   The B/P  shape in
this case  would be  sustained by  stars in  the so-called {\it banana} and
{\it anti-banana} orbit families, which  might be identified observationally
as asymmetries in the kinematics of the near and far side of the bulge
\citep[e.g.,][]{rangwala+2009, vasquez+2013}.

Following early studies by \citet{frogel+1987}, \citet{sharples+1990},
\citet{minniti+1996},  and  \citet{tiede+1997},  the  first  extensive
kinematical study of  the Galactic bulge is the  Bulge Radial Velocity
Assay     survey    \citep[BRAVA;][]{rich_brava+2007,     howard+2009,
  kunder+2012}, targetting $\sim 10,000$ M  giants from the Two Micron
All-Sky   Survey    \citep[2MASS;][]{skrutskie+2006},   at   latitudes
$b=-4,-6,-8$ and longitudes  $-10<l<10$. The main result  of the BRAVA
survey,  based  on  $\sim$4500  stars,   is  the  determination  of  a
cylindrical rotation pattern  for bulge stars, implying  that a simple
B/P bulge model  is sufficient to reproduce the  bulge kinematics with
no  need   for  a  merger-made  classical   bulge  \citep{howard+2009,
  shen+2010}. Dynamical models  have shown, however, that  a bar could
form also where a {\it classical} bulge (i.e.; formed via mergers) was
already  present.   In this  case,  then,  the classical  bulge  would
spin-up  to  a  faster  rotation   after  the  formation  of  the  bar
\citep{saha+2012} such that it would  be very difficult, a posteriori,
to detect its presence via kinematics alone \citep{gardner+2013}.

Independent clues on the bulge  formation mechanisms and timescale can
come   from    the   surface   chemical   abundance    of   individual
stars. Following several measurements based either on photometry or on
low  resolution  spectroscopy  \citep[e.g.,][]{rich+1988},  the  bulge
metallicity  distribution  function  was   first  derived  using  high
resolution spectra  by \citet{mcw-rich+1994}  and the  following works
obtained    consistent   results    \citep{ibata+1995a,   ibata+1995b,
  minniti+1996,      sadler+1996,     ramirez+2000,      zoccali+2003,
  fulbright+2006}.   All these  studies, however,  were confined  to a
single  low  reddening   window,  close  to  the   Baade's  Window  at
$(l,b)=(0,-4)$.    \citet{zoccali+2008},    and   \citet{johnson+2011,
  johnson+2013},  based   exclusively  on  high   resolution  spectra,
extended  previous  studies  to   a  few  additional  windows,  firmly
establishing the  presence of a  radial metallicity gradient  of $\sim
0.6$ dex per kpc,  with the most metal rich stars  being closer to the
Galactic center.  \citet{melendez+2008, alvesbrito+2010, johnson+2011,
  johnson+2013, gonzalez+2011_alphas}  found a similarity  between the
alpha  over  iron abundance  ratio  of  bulge  and thick  disk  stars,
suggestive of a fast formation  timescale for both components, perhaps
sharing a common origin \citep[see also][]{bensby+2011}.  By combining
[Fe/H] and  [Mg/Fe] abundances  and kinematics,  \citet{hill+2011} and
\citet{babusiaux+2010}   suggested  the   presence  of   two  distinct
components    in   the    galactic   bulge,    a   metal    poor   one
([Fe/H]$\sim$$-0.3$)  with  kinematics   typical  of  an  axisymmetric
spheroid, and  a metal rich one,  ([Fe/H]$\sim$+0.3) more concentrated
towards  the  Galactic plane,  with  a  significant vertex  deviation,
suggestive of a bar-like component.

A chemical and kinematical study covering a significantly larger bulge
area is  the Abundances  and Radial  velocity Galactic  Origins Survey
\citep[ARGOS;][]{freeman+2013, ness+2013a, ness+2013b}. The ARGOS team
measured  radial  velocities,  [Fe/H]  and  [$\alpha$/Fe]  ratios  for
$\sim$28,000 stars, $\sim$14,000 of which  at a distance of $<3.5$ kpc
from  the Galactic  center.  They  confirmed the  cylindrical rotation
found by  the BRAVA survey,  and could identify three  main components
within the bulge, tentatively associated with the metal rich B/P bulge
(mean     [Fe/H]$\approx$+0.15),      the     thick      B/P     bulge
([Fe/H]$\approx-0.25$)    and    the    inner   thick    disk    (mean
[Fe/H]$\approx-0.70$).  A different fraction  of the three populations
along different lines of sight gives rise to the observed gradients. A
first contiguous  map of the  mean metallicity  of the stars  over the
bulge  area  outside  $|b|$$\sim$2   has  been  derived  from  VVV
photometry by \cite{gonzalez+2013}.

The  GIBS  and  ARGOS  surveys  are  qualitatively  similar,  but  are
complementary with respect to the area coverage.  Our aim is to derive
Calcium II Triplet (CaT) metallicities,  radial velocities and, in the
near future,  proper motions  for a sample  of $\sim$5000  bulge stars
spread  across  the  area  covered by  the  VVV  survey  ($-10<l<+10$,
$-10<b<+5$).  We calibrate CaT equivalent  widths versus [Fe/H] with a
large ($\sim$400) sample of stars  observed both with intermediate and
high resolution  spectroscopy. In  the present  paper we  describe the
target,  selection, observations  and we  provide the  radial velocity
catalogue for all our targets.

%___________________________________________________________________
\section{Observations}

%________________
\subsection{Target selection}

\begin{figure}[h]
\centering
\includegraphics[width=9cm, angle=0]{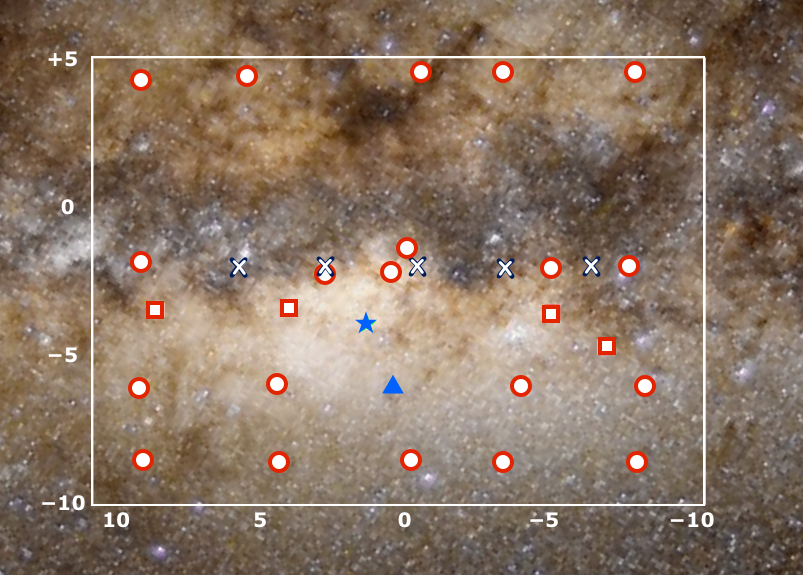}
\caption{The  approximate  location of  the  fields  discussed in  the
  present  paper, overplotted  on an  optical image  of the  Milky Way
  bulge (\copyright Serge Brunier). The  large white rectangle is the
  area  mapped by  the  VVV  survey.  Red and white circles are  fields
  observed at  low spectral  resolution (R$=6500$) through  setup LR8,
  from programme 187.B-0909; red and white squares  are fields observed at high
  resolution   (R$=22500$)   through   setup  HR13,   from   programme
  187.B-0909.  Crosses at $b$$\sim$$-2$ are  the fields from programme 089.B-0830,
  also  observed through  LR8.  The  blue star  is  the field,  within
  Baade's Window, used for the  CaT calibration, and the blue triangle
  is the field discussed in \cite{vasquez+2013}.}
\label{VVVmap}
\end{figure}

\begin{figure}[h]
\centering
\includegraphics[width=9cm, angle=0]{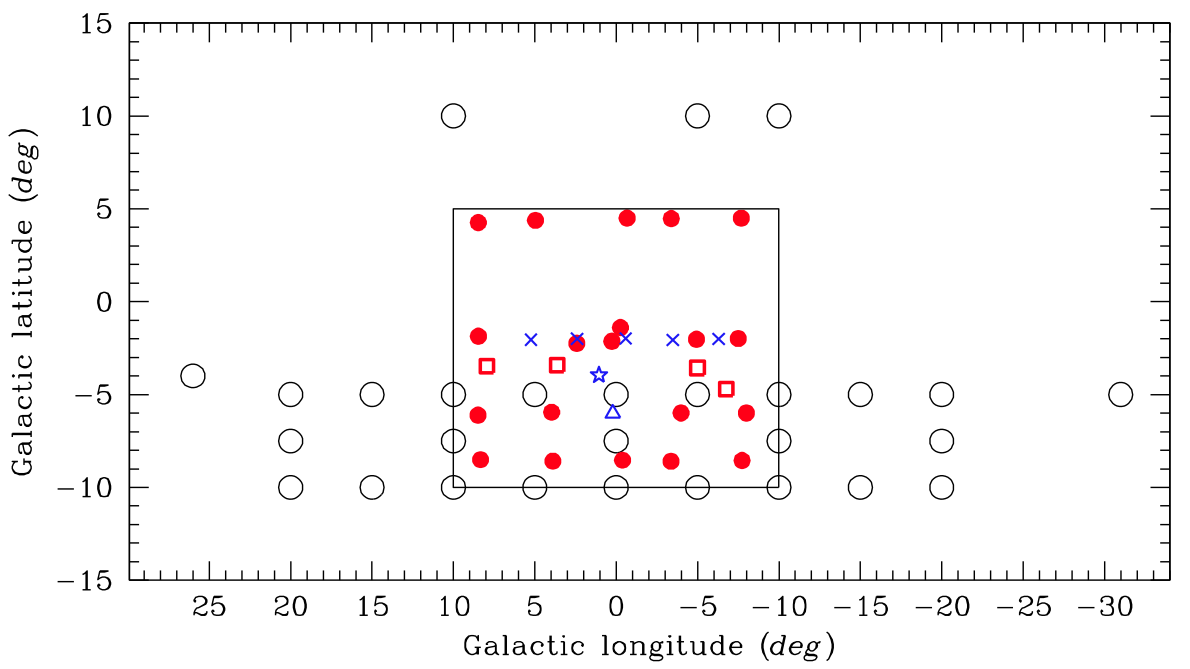}
\caption{The location  of the  fields discussed  in the  present study
  compared    to     the    fields     from    the     ARGOS    survey
  \citep{freeman+2013}. Black  open circles are the  ARGOS fields. Red
  filled  circles  are  fields  observed at  low  spectral  resolution
  (R$=6500$)  through  setup  LR8,  from  programme  187.B-0909;  open
  squares are  fields observed at high  resolution (R$=22500$) through
  setup HR13, from programme 187.B-0909.   Blue crosses are the fields
  from programme 089.B-0830, also observed through LR8.  The blue star
  is the field,  within Baade's Window, used for  the CaT calibration,
  and the blue triangle is  the field discussed in \cite{vasquez+2013}
  . The large rectangle shows the area mapped by the VVV survey.}
\label{argos}
\end{figure}

Target stars for the spectroscopic observations were selected from the
VVV catalogues. The catalogues contain aperture magnitudes for individual
stars in the  $J,H,K_s$ bands derived from  the Cambridge Astronomical
Survey Unit  (CASU) pipeline,  and have  been complemented  with 2MASS
data \citep{cutri03} for stars brighter  than $K_s$=12, where VVV images
are saturated.  Further  details on the photometric  catalogues can be
found in \cite{gonzalez+2011redd, saito+2012}.

The goal  of the programme  was to collect spectra  for representative
samples of  bulge field  K giants  spread over a  grid, as  regular as
possible, across the  bulge area covered by the  VVV survey.  However,
starting  from  a  regular  position grid  at  $b=-8,-6,-4,-2,+4$  and
$l=-8,-4,0,+4,+8$, the actual  center of each field  (see Table~1) was
fine-tuned on  the extinction  map derived by  \cite{gonzalez+2012} in
order to minimize  the reddening, hence maximizing the S/N  at a given
exposure time.  Occasionally it  was also  moved by  up to  $1\deg$ in
order to  overlap with previous photometric  observations.  The latter
were  used to  add  extra information  on the  target  stars, such  as
photometry  in different  bands  and proper  motions  from the  OGLEII
survey  \citep{sumi04pm}.    An  additional  field  was   observed  at
($l,b$)=($0,-1$)   in  order   to  investigate   whether  the   radial
metallicity gradient seen along  the minor axis by \cite{zoccali+2008}
and \cite{johnson+2011,johnson+2013}  extends to the inner  regions or
flattens out as suggested by \cite{rich07, rich+2012}.

Three aditional sets of spectra were added to the GIBS sample in order
to obtain a  finer field grid in  the sky.  The first  one consists of
LR08  spectra  in  5  fields  at latitude  $b=-2$,  from  ESO  program
089.B-0830 (PI. Gonzalez).   These data were obtained with  the aim of
characterizing   the  properties   of  the   inner  Bulge,   following
\cite{gonzalez+2011_bar}. The targets for this programme were selected
with  the same  criteria explained  below.   The location  of these  5
fields   is    shown   with    crosses   in    Fig.~\ref{VVVmap}   and
Fig.~\ref{argos}.   The second  set  consists of  spectra  for 111  RC
stars, in Baade's Window, obtained to derive the CaT calibration, used
to derive  metallicities for the  LR8 target stars.  Those  stars were
observed  through setup  LR8, but  they were  also observed  at higher
spectral     resolution     within      our     previous     programme
\citep{zoccali+2008}. Finally, in what follows  we will include in all
plots also the RC stars, at ($0,-6$), analysed in \cite{vasquez+2013}.

Figure~\ref{VVVmap} shows  the location of  the 31 fields  observed in
the present  study, at different spectral  resolutions, overplotted on
the  stellar  density  map from  \cite{saito84M}.   Figure~\ref{argos}
shows  the location  of  the same  fields  shown in  Fig.~\ref{VVVmap}
together   with  the   fields   observed  within   the  ARGOS   Survey
\citep{freeman+2013}.  The  two surveys nicely complement  each other,
in terms of  distance from the Galactic center.  Indeed,  only four of
the ARGOS fields are included within the area surveyed here.

\begin{figure}[h]
\centering
\includegraphics[width=9.5cm, angle=0]{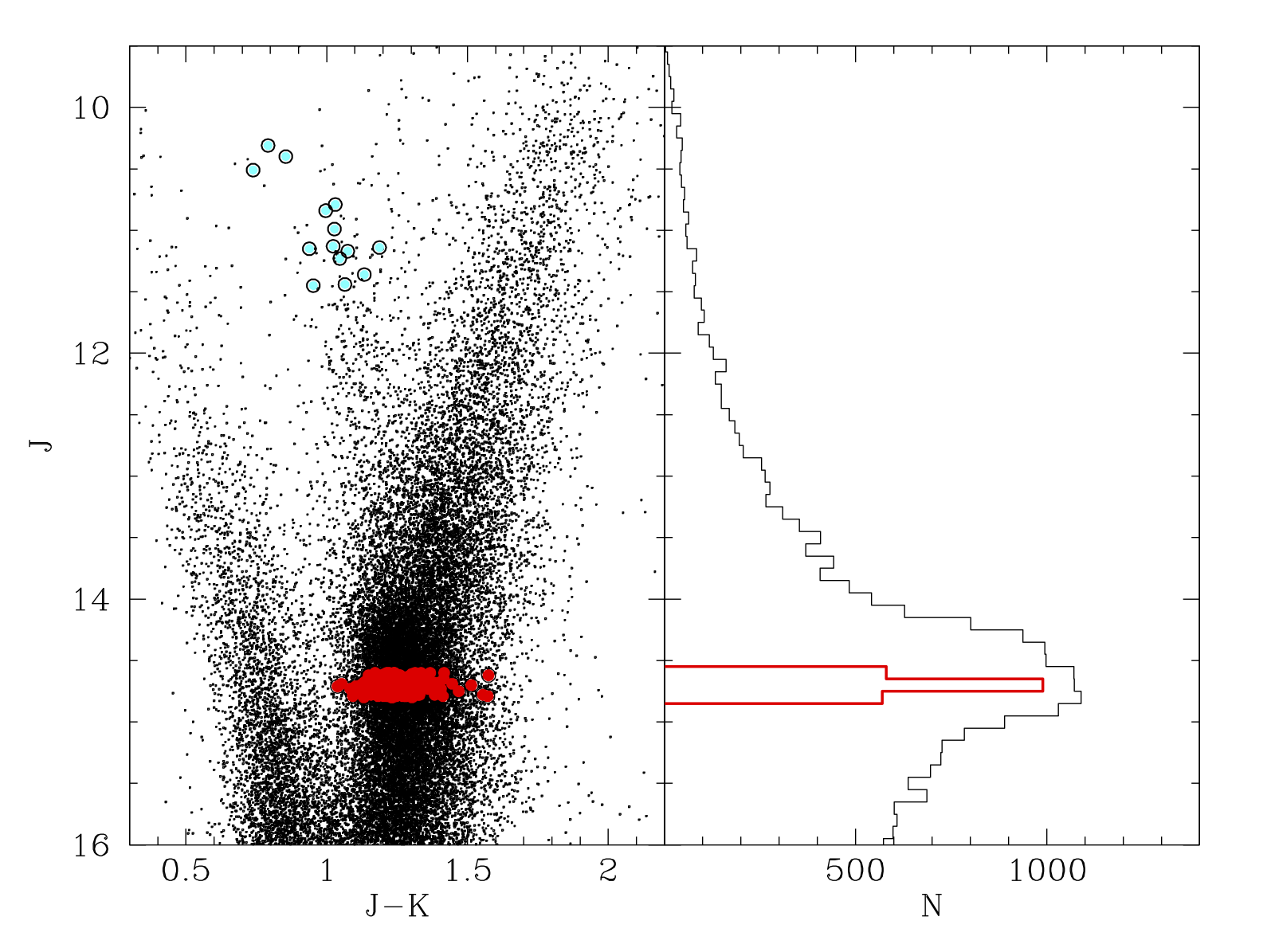}
\caption{Example of the GIBS target selection criteria for the LRm5m2 field.
{\it Left:} the VVV CMD together with the 214 GIRAFFE targets (red) and 
the 14 UVES targets (cyan). {\it Right:} magnitude distribution of the
GIRAFFE targets (red histogram) compared with the underlying luminosity 
function of the RGB stars with $J-K>1.0$ (black histogram).}
\label{targets}
\end{figure}

For each field,  after applying the reddening  correction, we selected
the most  likely bulge  members based  on the  red giant  branch (RGB)
color redder  being than $(J-K)_0\sim0.4$,  where the exact  limit was
adjusted slightly from  field to field.  This  cut excluded foreground
disk dwarfs,  without loosing  the blue  edge of  the bulge  red clump
(RC).  The  RGB luminosity function  was then constructed in  order to
identify the RC.  Target stars were  selected in a narrow range of $J$
magnitude (or,  whenever available, in $I$)  close to the RC  peak, as
shown  in Fig.~\ref{targets},  in order  to ensure  that the  acquired
spectra would have  similar signal-to-noise (S/N).  In  order to avoid
metallicity biases, possibly due to  color selection, we sampled stars
spanning the whole  color range of the  RC.  Figure~\ref{targets} also
shows the nearby disk RC stars  that have been observed in addition to
the  bulge  giants,  in  some fields,  using  the  simultaneous  fibre
connection to  the UVES  spectrograph (c.f.,  Table~\ref{log}).  These
UVES spectra will be the subject of a dedicated paper, and will not be
further discussed here.

% 4982 stars with LR8    R~6500   8206-9400 A
%  448 stars with HR13   R~22500  6120-6405 A
% UVES: 84 stars with Red580  R~45000  4800-6800 A

%________________
\subsection{Spectra}

\begin{table*}
\caption{Observed fields, number of target stars and main characteristics of the spectra.}
\label{log}
\centering
{\small
\begin{tabular}{l c c r r c c c r r r r}
\hline\hline
Field name & RA & DEC & l & b & Setup & R=$\lambda/\Delta \lambda$ & $\lambda$ coverage & N$_{\rm stars}$ & N$_{\rm stars}$ & exptime/star \\ % & S/N \\  
       & (hr) & (deg) & (deg) & (deg) & &                          &   $\AA$            & GIRAFFE       &  UVES          & (s) \\ % & \\       
\hline
LRp8p4 & 17:48:49.2 & -19:29:23.90 &   8.4712 &  4.2582 & LR8  &  6500 & 8206-9400 & 209 & -  & 3200  \\ % & \\
LRp5p4 & 17:40:35.5 & -22:25:04.20 &   4.9616 &  4.3818 & "    &   "   &    "      & 208 & -  & 3200  \\ % & \\
LRp0p4 & 17:26:47.5 & -27:04:35.00 & 359.3328 &  4.5033 & "    &   "   &    "      & 210 & -  & 13500 \\ % & \\
LRm3p4 & 17:20:00.0 & -29:20:00.00 & 356.6168 &  4.4750 & "    &   "   &    "      & 209 & -  & 2700  \\ % & \\
LRm8p4 & 17:08:15.3 & -32:48:34.80 & 352.3136 &  4.4984 & "    &   "   &    "      & 210 & -  & 2600  \\ % & \\
LRp0m1 & 17:50:28.7 & -29:52:43.40 & 359.7396 & -1.3930 & "    &   "   &    "      & 441 & 14 & 13500 \\ % & \\
LRp8m2 & 18:11:35.0 & -22:31:43.80 &   8.4699 & -1.8609 & "    &   "   &    "      & 209 & 10 & 9600  \\ % & \\
LRp3m2 & 18:00:01.0 & -27:59:22.00 &   2.4243 & -2.2435 & "    &   "   &    "      & 207 &  6 & 4000  \\ % & \\
LRp0m2 & 17:54:38.9 & -29:48:01.80 &   0.2668 & -2.1318 & "    &   "   &    "      & 435 &  7 & 7800  \\ % & \\
LRm5m2 & 17:41:34.5 & -34:11:35.60 & 355.0712 & -2.0236 & "    &   "   &    "      & 209 & 14 & 8100  \\ % & \\
LRm8m2 & 17:34:41.2 & -36:20:48.40 & 352.4983 & -1.9858 & "    &   "   &    "      & 210 &  7 & 13500 \\ % & \\
HRp8m3 & 18:16:40.8 & -23:45:32.20 &   7.9460 & -3.4770 & HR13 & 22500 & 6120-6405 & 106 &  7 & 27000 \\ % & \\
HRp4m3 & 18:07:15.4 & -27:31:21.70 &   3.6174 & -3.4111 &  "   &  "    &     "     & 91  &  5 & 27000 \\ % & \\
HRm5m3 & 17:47:49.2 & -35:03:24.10 & 355.0036 & -3.5701 &  "   &  "    &     "     & 108 &  7 & 27000 \\ % & \\
HRm7m4 & 17:48:11.0 & -37:09:25.30 & 353.2336 & -4.7106 &  "   &  "    &     "     & 108 &  7 & 27000 \\ % & \\
LRp8m6 & 18:28:00.0 & -24:30:00.00 &   8.4894 & -6.1089 & LR8  &  6500 & 8206-9400 & 209 & -  & 1700 \\ % & \\
LRp4m6 & 18:18:08.0 & -28:25:20.00 &   3.9663 & -5.9517 & "    &   "   &    "      & 213 & -  & 1500 \\ % & \\
LRm4m6 & 18:00:34.0 & -35:23:42.00 & 356.0189 & -5.9922 & "    &   "   &    "      & 224 & -  & 1500 \\ % & \\
LRm8m6 & 17:50:38.4 & -38:51:49.00 & 352.0041 & -5.9944 & "    &   "   &    "      & 217 & -  & 2500 \\ % & \\
LRp8m8 & 18:37:09.2 & -25:42:57.30 &   8.3281 & -8.5060 & "    &   "   &    "      & 194 & -  & 1800 \\ % & \\
LRp4m8 & 18:28:41.8 & -29:41:28.00 &   3.8859 & -8.5847 & "    &   "   &    "      & 208 & -  & 1800 \\ % & \\
LRp0m8 & 18:19:34.8 & -33:26:31.80 & 359.6151 & -8.5338 & "    &   "   &    "      & 417 & -  & 2100 \\ % & \\
LRm3m8 & 18:13:15.7 & -36:05:51.00 & 356.6270 & -8.5928 & "    &   "   &    "      & 208 & -  & 1500 \\ % & \\
LRm8m8 & 18:02:49.7 & -39:53:14.00 & 352.2678 & -8.5491 & "    &   "   &    "      & 207 & -  & 1500 \\ % & \\
       &            &              &          &         &      &       &           &          & \\ % & \\
\multicolumn{4}{l}{~~ Additional archive fields:} \\
LRp5m2-OG & 18:05:27.31   & -25:26:51.5  &    5.2366 &  -2.0500 & "    &   "   &    "      & 112 & -  & 1200 \\ % & \\
LRp3m2-OG & 17:59:00.73   & -27:53:40.6  &    2.3964 &  -2.0039 & "    &   "   &    "      & 113 & -  & 1200 \\ % & \\
LRm1m2-OG & 17:52:03.13   & -30:27:15.9  &  359.4181 &  -1.9794 & "    &   "   &    "      & 111 & -  & 1200 \\ % & \\
LRm3m2-OG & 17:45:21.10   & -32:59:02.2  &  356.5143 &  -2.0598 & "    &   "   &    "      & 111 & -  & 1200 \\ % & \\
LRm6m2-OG & 17:37:59.13   & -35:20:15.4  &  353.7093 &  -2.0068 & "    &   "   &    "      & 113 & -  & 1200 \\ % & \\
Baade's Win & 18:04:51.19 & -30:03:26.6  &    1.1400 &  -4.1800 & "    &   "   &    "      & 111 & -  & 1200 \\ % & \\ 
LRp0m6$^{(a)}$    & 18:10:18.33   & -31:45:11.8  &    0.2100 &  -6.0200 & --   &  --   &   --      & 454 & -  & 2700/1800 \\ % & \\
       &            &              &          &         &      &       &           &          & \\ % & \\
\multicolumn{11}{l}{$^{(a)}$ These are the spectra, from GIRAFFE@VLT + IMACS@Magellan, discussed in \cite{vasquez+2013}. } \\
\hline
\end{tabular}
}
\end{table*}

\begin{figure}[h]
\centering
\includegraphics[width=9cm, angle=0]{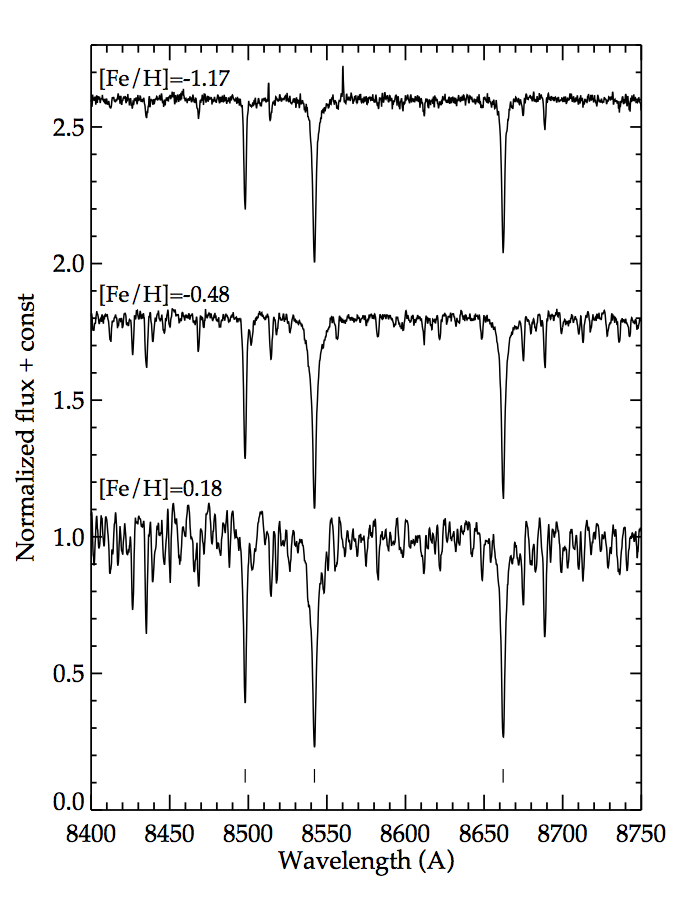}
\caption{Example  of typical  spectra observed  through setup  LR8 for
  three  targets with  different  preliminary metallicities.  Vertical
  ticks mark the three corresponding CaT lines.}
\label{spec}
\end{figure}

Spectra for the selected targets  have been collected with the GIRAFFE
spectrograph fed by MEDUSA fibres (in some cases the observations were
taken  in the  combined UVES+MEDUSA  mode) of  the FLAMES  multi-fibre
instrument  \citep{pasquini+2002} at  the  ESO  Very Large  Telescope,
between May 2011 and September 2012.  Observations were carried out in
service  mode under  programme  187.B-0909(A)  and 187.B-0909(B),  PI:
Zoccali.  The log of the  observations is reported in Table~\ref{log}.
All  the  spectra, with  the  exception  of  those  in the  fields  at
$b$$\sim$$-4$,  were  obtained  with setup  LR8,  at  resolution
R=6500,  centered  on   the  Calcium  II  Triplet   (CaT)  feature  at
$\sim$8500$\AA$.   Two  different  sets   of  fibre  allocations  were
performed  for  each field,  allocating  up  to 132$\times$1.2  arcsec
diameter fibres in  each setup. Approximately 25 fibres  in each field
were allocated to  empty sky positions, thus collecting  spectra for a
total of  $\sim$215 targets  per field.  Four  independent allocations
were  done  for the  fields  along  the  minor axis,  thus  collecting
$\sim$450  targets   in  these  fields.   Indeed,   these  fields  are
particularly interesting because  they cross two arms  of the X-shaped
bulge,  at  latitude  $|b|$>5.   For  $|b|$<3  the  larger
statistics was  motivated by the  need to solve the  debate concerning
the extension  of the  metallicity gradient to  the inner  bulge.  The
exposure time was optimized to reach  a S/N$\sim$50 per pixel.  In the
most  reddened  fields  this   required  multiple  exposures.  Typical
spectra,  for three  stars of  different metallicities,  are shown  in
Fig.~\ref{spec}.

Stars in  the fields  at $b$$\sim$$-4$  were observed  at higher
resolution (R=22,500) through setup HR13 centered at $\sim 6300\AA$ in
order  to  measure  the  chemical  abundance of  iron  and  the  light
elements, and to  investigate the presence of radial  gradients in the
alpha element ratios across different  longitudes. The typical S/N per
pixel of these HR spectra is $\sim$100.

%___________________________________________________________________
\section{Data Reduction Pipeline}

In this  section we describe  our full pipeline,  including extraction
and calibration  of the spectra  as well  as the adopted  procedure to
obtain measurements  of physical parameters from  the reduced spectra.
Only the  radial velocitites  are discussed in  detail in  this paper,
while  the  analysis  of  the  metallicity  and  individual  elemental
abundances  will   be  discussed  in  forthcoming,   dedicated  papers
(V\'asquez et al 2014, Gonzalez et al. 2014, in preparation).

The spectra were extracted and wavelength calibrated using the GIRAFFE
pipeline maintained by ESO, which processes the spectra applying bias,
flat-field correction, individual  spectral extraction, and wavelength
calibration based  on daytime  calibration frames. Since  the pipeline
does not perform sky subtraction,  the correction was done using IRAF
tasks. As a first step, a master  sky has been obtained for each field
from the ~20  sky spectra, median combined and using  a sigma clipping
algorithm. The master  sky was then subtracted from  the 1--D spectrum
of each target using the  IRAF \textit{skytweak} task, shifting and/or
scaling  the input  sky  spectra  to improve  the  subtraction of  sky
features from target spectra.

\subsection{Radial velocities}

Heliocentric radial  velocities where measured  by cross--correlations
using IRAF  \textit{fxcor} task.  For  the low resolution  spectra the
adopted  template  was  a   synthetic  spectrum,  generated  with  the
Turbospectrum  code \citep{alvarez+1998},  fed  with  the MARCS  model
atmosphere  \citep{gustafsson+2008},  for stellar  surface  parameters
appropriate  for   a  metal  poor  bulge   K  giant  ($T_{eff}=4750$K,
log$g=2.5$ and [Fe/H]$=-1.3$).  The template metallicity was chosen on
the low side of the distribution  in order to avoid including too many
small lines  that would add  only noise in the  cross-correlation peak
for metal-poor  stars.  The same  template was  used for stars  of all
metallicities. The template  covers the CaT region from  8350$\AA$ to
8950$\AA$.   The statistical  error on the  radial velocity  from low
resolution spectra is typically $\sim  1$ km/s. The final heliocentric
radial velocities for  all the target stars observed both  at high and
low spectral resolution are listed in Table~\ref{tab_RVs}.

For  the high  resolution,  HR13 spectra,  the  cross correlation  was
performed with the same IRAF routine but a template synthetic spectrum
with  $T_{eff}=4500$K, log$g=2.3$  and [Fe/H]$=-0.3$.  The latter  was
generated  adopting the  MARCS  model atmospheres  and  the MOOG  code
--version 2010--  for spectrum synthesis \citep{sneden+1973}.  In this
case there were multiple exposures  ($\sim$10) for each field in order
to  reach the  required S/N.   The cross  correlation was  carried out
independently on  each individual exposure  and the results  were then
averaged    to     the    final    radial    velocity     listed    in
Table~\ref{tab_RVs}.  The   typical  error  on  these   velocities  is
$\sim$0.6 km/s,  calculated as  the standard  deviation of  the radial
velocity distribution from individual exposures.

\subsection{Chemical abundances}

In forthcoming  articles, we  will present a  catalog and  analyze the
[Fe/H]  measurements  for  all  the stars  observed  at  low  spectral
resolution (Vásquez et  al. 2014a, in preparation), as  well as [Fe/H]
and element  ratios (mainly  [alpha/Fe]) for  targets observed  at the
high  resolution   using  HR13  setup   (Gonzalez  et  al.   2014,  in
preparation). Hereafter we briefly describe our pipeline and procedure
adopted to derive chemical abundance measurements.

\subsubsection{Low resolution spectra}

Iron  abundances from  LR  spectra  are obtained  using  CaT lines  as
metallicity indicator.  The correlation  between the equivalent widths
of  CaT  lines  and  global  metallicity  was  first  demonstrated  by
\cite{armandroff+1988}  by means  of  integrated  spectra of  Galactic
globular clusters. Later  on this empirical evidence  was confirmed in
several  studies  of   individual  star  spectra,  and   it  has  been
extensively used in the study of  Galactic star clusters and Milky Way
satellites   \citep[e.g.,][,  and   references  therein]{carrera+2007,
  battaglia+2008,  starkenburg+2010,  saviane+2012}.  Before  starting
the present programme we made sure  that such a correlation would hold
for super  solar metallicities,  with the [Ca/Fe]  profile appropriate
for  bulge K  giants.  Specifically,  we observed  a set  of $\sim$200
bulge RC and red giant branch stars in Baade's Window both through the
low  resolution, LR8,  CaT setup  and through  three setups  at higher
resolution \citep{zoccali+2008,  hill+2011}.  These  observations were
used to derive  a CaT versus [Fe/H]  calibration specifically designed
for  the targets  of  the  GIBS Programme.   The  calibration will  be
presented and discussed in V\'asquez et al.  (2014b, in preparation).

\subsubsection{High resolution spectra}

For  the stars  observed  at R$\sim$22,500,  through  the HR13  setup,
metallicities and element ratios  are derived using the same
iterative    method    described     in    \citet{zoccali+2008}    and
\citet{gonzalez+2011_alphas}.   Specifically,   equivalent  widths  of
isolated  Fe lines  are  obtained automatically  by  means of  DAOSPEC
\citep{daospec}.   A  first  guess  photometric  temperature  will  be
derived  using the  $(V-I)$ colors  from OGLEII  \citep{udalski+2002},
de-reddened  based  on  the   high  resolution  extinction  maps  from
\citet{gonzalez+2012}.   The  effective   temperature  calibration  by
\cite{alonso+1999}  are   used.   The  same  work   yields  bolometric
corrections, used to  estimate photometric gravities, by  means of the
classical formula:

\begin{footnotesize}
\[
\log\left(g\right)=\log\left(g_{\odot}\right)+\log\left(\frac{M_*}{M_{\odot}}
\right)+0.4\left(M_{Bol,*}-M_{Bol,\odot}\right)+4\log\left(\frac{T_{eff,*}}{T_{eff,\odot}}\right)
\]
\end{footnotesize}

where   $\rm   M_{Bol,\odot}=4.72$,  $\rm   T_{eff,\odot}=5770$K   and
$\log\left(g_{\odot}\right)=4.44$dex. Absolute  visual magnitudes were
obtained  assuming   distances  to  each   field  from  the   work  of
\citet{gonzalez+2013}.  Microturbulence  velocity and  global metallicity
are set to  1.5 and 0.0, respectively, as a  first step.  These values
are used  to obtain a first  guest MARCS stellar model  atmosphere and
are subsequently  refined spectroscopically.   Spectroscopic effective
temperatures   and   microturbulence   velocities   are   refined   by
simultaneously requiring  excitation equilibrium  and a flat  trend of
iron abundance versus equivalent  width, respectively. A new iteration
is then started  by re-calculating the {\it  photometric} gravity with
the new  temperature and  iron abundance, the  latter entering  in the
derivation of the bolometric correction.

Alpha  element abundances  are  derived using  the  final MARCS  model
atmosphere for  each target  star to  produce synthetic  spectra using
MOOG. These synthetic spectra are  compared with the observed ones and
abundances for  Mg, Ca, Si  and Ti are  then derived by  fitting their
corresponding spectral lines in our spectral range.

%___________________________________________________________________
\section{Radial Velocities}
    
\begin{table*}
\caption{Coordinates and radial velocities for all the program stars.}
\label{tab_RVs}
\centering
{\small
\begin{tabular}{l r c c c c c c}
\hline\hline
Field name & star ID & RA & DEC & l & b & $V_{\rm R}$ & $\sigma V_{\rm R}$ \\
           &         & (hr) & (deg) & (deg) & (deg) & (km/s) & (km/s) \\
\hline
LRp0m1 & GIBS\_1	& 17:50:19.76 & -29:44:46.7 & 359.83672 & -1.29727 & 181.5 & 1.7 \\
LRp0m1 & GIBS\_2	& 17:50:26.67 & -29:52:28.7 & 359.73928 & -1.38447 & 107.7 & 1.8 \\
... & ... & ... & ... & ... & ... & ... & ... \\
\\
\hline
\end{tabular}
}
\end{table*}

\begin{figure*}[!ht]
\centering
\includegraphics[width=16cm, angle=0]{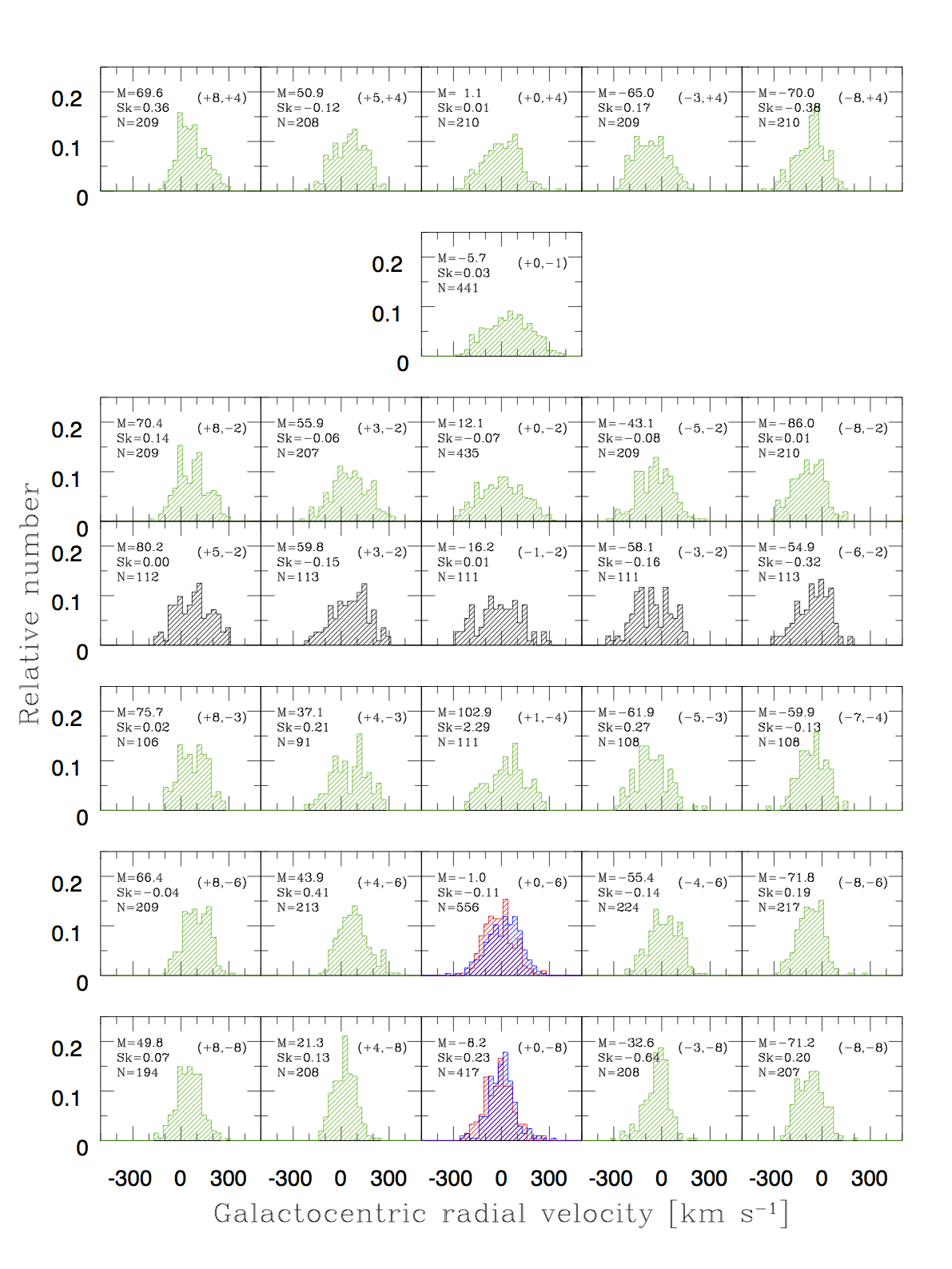}
\caption{Radial velocity  distributions for  all the GIBS  fields. The
  two fields  at ($0,-6$)  and ($0,-8$) include  targets both  for the
  bright RC  (red) and  the faint  (RC). Part of  the data  at $b=-2$,
  shown  as  black histograms,  come  from  programme 089.B-0830.  The
  labels  on the  upper  left corner,  in each  panel,  list the  mean
  velocity (M), the skewness (Sk) and  the number of stars (N). On the
  upper right corner we list the galactic coordinates of the field.  }
\label{rv}
\end{figure*}

The radial velocity distributions for the observed fields are shown in
Fig.~\ref{rv}. The  overall distributions  are in good  agreement with
those  observed  from the  BRAVA  survey  in the  overlapping  regions
($b<-$5).   Our  survey  allows  us to  investigate  the  radial
velocity  distribution  of  the  inner bulge  regions  for  the  first
time. Velocities  range from $-300$ to  $+300$ km/s with a  shape that
changes considerably across the bulge  area. It is worth noticing that
no significant peaks  (nor individual outliers) are  found outside the
main   distribution,  in   contrast  with   the  findings   by
\citet[e.g.,][]{nidever+2012}.

In order to search for the  presence of systematic trends in the shape
of the  distributions across the  different fields, we  calculated the
skewness  of the  radial velocity  distribution in  each field.  These
values are  also shown in  Fig.~\ref{rv}. With the exception  of field
HRp1m4 and  LRm3m8, which show  a significative positive  and negative
skewness  respectively, all  fields  are consistent  with a  symmetric
distribution. This indicates that  the mean radial velocity variations
across  the  different  fields  are  most  likely  due  to  a  general
\textit{shift}  of the  distribution instead  of being  the result  of
additional features such as asymmetric tails or local peaks.

\subsection{Rotation curves and cylindrical rotation}

The observed rotational  profiles of bulges can be  directly linked to
the different  processes involved  in the  formation and  evolution of
bulges.  In particular, bulges that  originate from the buckling
instability  of  a,  previously  settled, rotating  bar,  when  viewed
edge-on are  expected to  show little difference in  their mean
rotation velocities measured at different scale heights from the plane
of  the galaxy.  This is  the  well known  property named  cylindrical
rotation.  For the  Milky Way, \citet{shen+2010} compared  models of a
secularly evolved bar with the observed rotation curve obtained within
the  BRAVA survey.   They  concluded that  a model  with  a pure  disk
component was sufficient to explain the observed rotation, without the
need to add a merger made, {\it classical} bulge.

\begin{figure}[h]
\centering
\includegraphics[width=9cm, angle=0]{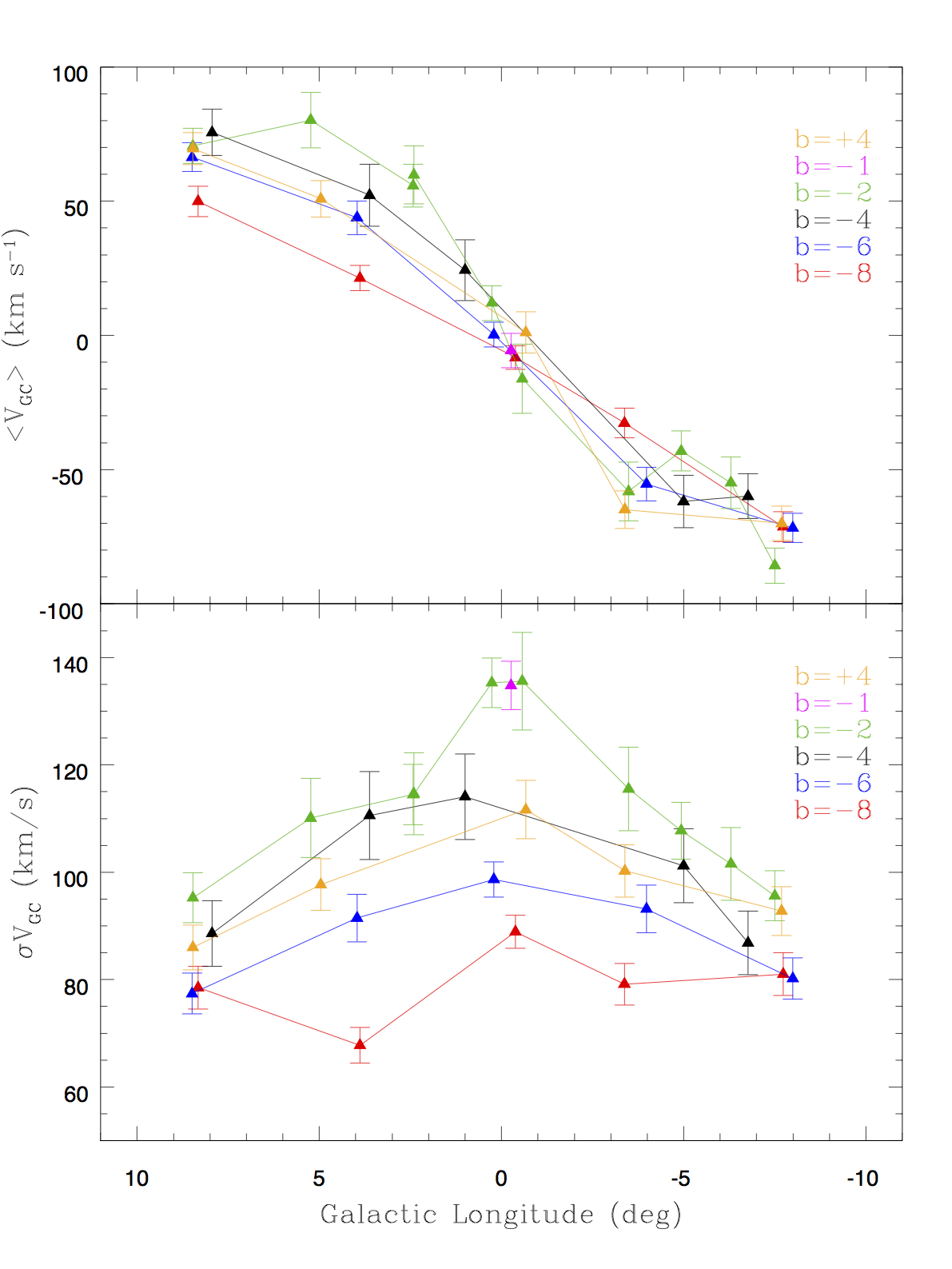}
\caption{Mean  galactocentric  radial   velocity  (top)  and  velocity
  dispersion  (bottom)  as  a  function  of  Galactic  longitude,  for
  different latitudes, as listed in the labels.}
\label{rotation}
\end{figure}

Mean radial velocities  for each of our fields were  computed and used
to construct  the rotational profiles at  different vertical distances
from the  Galactic plane.  Figure~\ref{rotation} (top)  shows the mean
radial velocity, corrected  for the motion of the Sun  with respect to
the  Galactic   center using the formula \cite[e.g.,][]{ness+2013b}:
\[
V_{\rm GC}=V_{\rm HC}+220\sin(l)\cos(b)+16.5[\sin(b)\sin(25) + \cos(b)\cos(25)\cos(l-53)]
\]

\begin{figure}[h]
\centering
\includegraphics[width=9cm, angle=0]{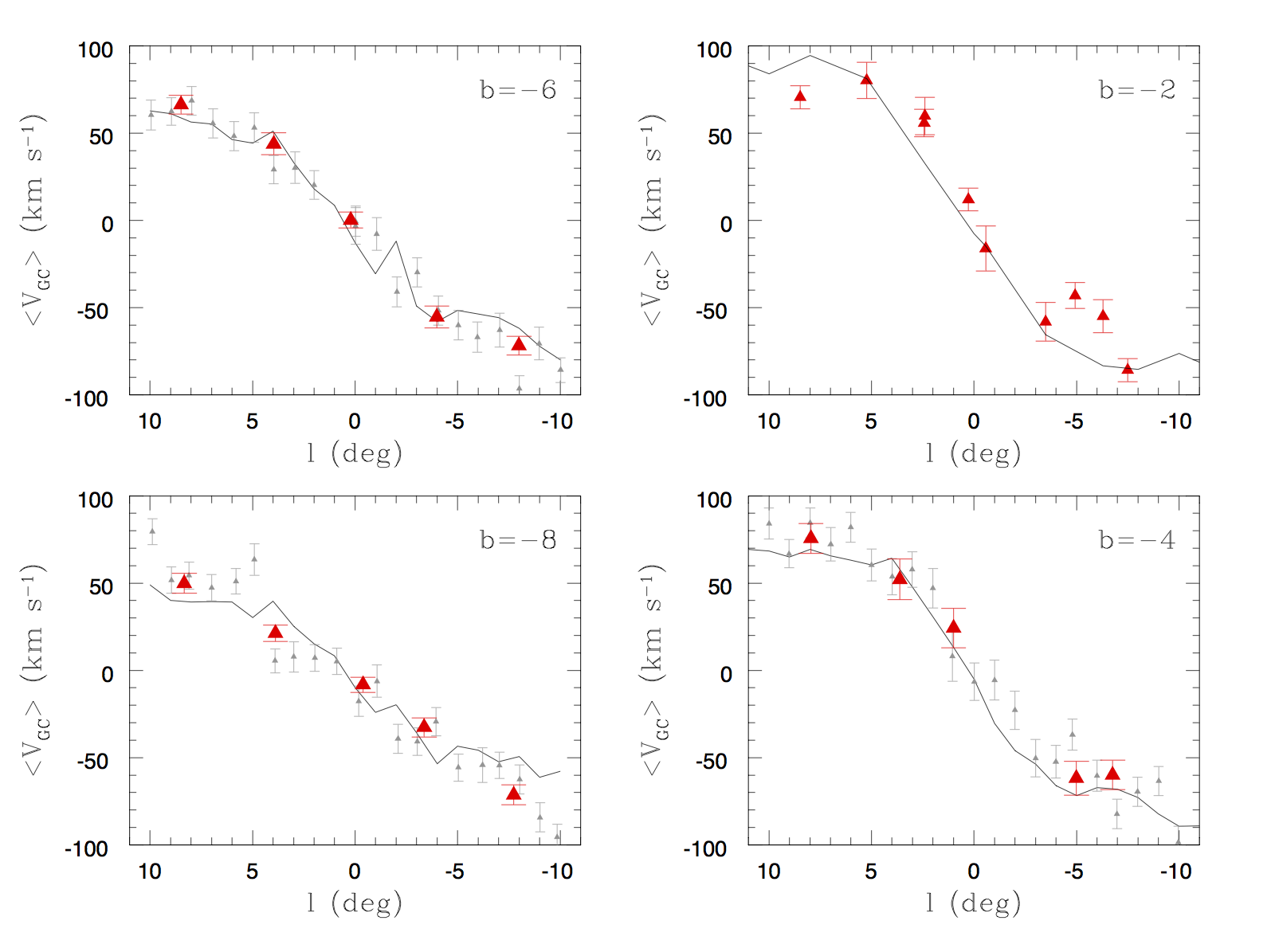}
\caption{Mean galactocentric radial velocity as a function of Galactic
  longitude  for GIBS  fields at  different latitudes  (red triangles)
  compared  to the  models by  \cite[][solid lines]{inma+2006}  at the
  same latitudes. Data from the BRAVA survey, when available, are also
  plotted in gray.}
\label{model_rv}
\end{figure}

\begin{figure}[h]
\centering
\includegraphics[width=9cm, angle=0]{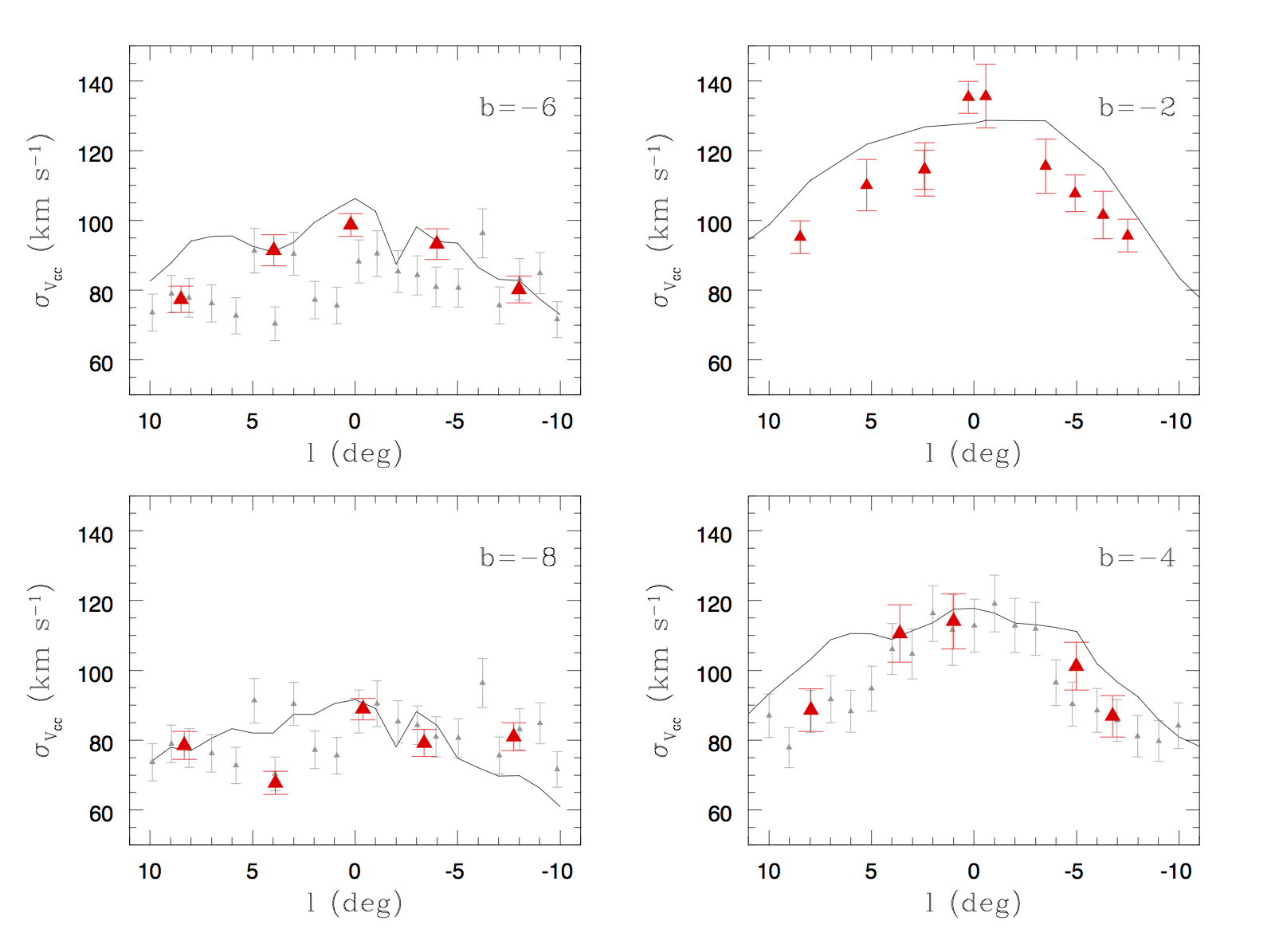}
\caption{Radial  velocity   dispersion  as  a  function   of  Galactic
  longitude  for GIBS  fields at  different latitudes  (red triangles)
  compared  to the  models by  \cite[][solid lines]{inma+2006}  at the
  same latitudes. Data from the BRAVA survey, when available, are also
  plotted in gray.}
\label{model_sig}
\end{figure}

We will refer to $V_{\rm GC}$ as {\it galactocentric radial velocity}
hereafter. The lower  panel of the
same  figure shows  the radial  velocity dispersion  as a  function of
longitude, for different latitudes. It  is clear from this figure that
the  radial velocity  curve,  reflecting the  bulge rotation,  becomes
steeper  towards lower  latitudes.   This effect,  already noticed  by
\citet{howard+2009}, is  now clearly  confirmed with the  inclusion of
the rotation  profile at  $b=-2$.  As discussed  in \citet{inma+2013},
this is the expected  rotation pattern of a B/P bulge  formed out of a
bar with  a non-zero position  angle with respect to  the Sun-Galactic
centre  line of  sight. No further component  needs to  be included
(although  neither necessarily  excluded) to  reproduce the  observed
rotation  curves.  Indeed,  a direct  comparison between  our rotation
profiles with those of the  B/P bulge model based on \citet{inma+2006}
and   presented   in    \citet{inma+2011}   (Fig.~\ref{model_rv}   and
Fig.~\ref{model_sig}) shows a remarkable agreement at all latitudes.

In the model from \citet{inma+2006} a  boxy bulge is formed out of the
buckling instability  of a  bar with no  inclusion of  any merger-made
component.  This  model  successfully  reproduces  a  number  of
  observations of the  Milky Way bulge, including the  RC star counts
  in the innermost region \citep{gerhard+2012}, the  RC splitting at
  l=0,  |b|>5, the metallicity profile across ($l,b$) and the observations
  of the apparent change in the inclination angle of the bar outside 
  the boxy/peanut extent, at $l\sim$27 \citep{inma+2011}.
Stars in
the models were  selected within a distance  corresponding to $\pm0.2$
magnitudes, around the bar major axis.  In Fig.~\ref{model_rv} we show
how it reproduces the rotation profile of the Bulge, as found with the
GIBS (red) and the BRAVA (gray) data.  The very good agreement between
this  model  and  the  data   supports  the  conclusion  presented  in
\citet{shen+2010}, extending it to the inner bulge at $b=-2$.

It should  be mentioned that a  few points at $l<-3$,  $b=-2$ do stick
out of the main observational  trend, and show a galactocentric radial
velocity significantly higher than predicted  by the model. Keeping in
mind that these points correspond to  the far side of the bar, sampled
very close to  the Galactic plane, we believe that  the discrepancy is
likely  due to  the fact  that  the line  of sight  in that  direction
samples stars in the  near half of the bar more than  stars on the far
half, due to a simple projection effect.  This well known effect has been
discussed by, e.g., \cite[][(their Appendix A)]{lopez-corredoira+2007}
to explain  why the maximum density  position along the line  of sight
does not coincide with the intersection between the bar major axis and
the line of sight. The effect is  present in the data more than in the
models because of the different selection criteria, explained above.

Figure~\ref{model_sig} compares  the velocity dispersion  between GIBS
(red) and BRAVA  (gray) samples as well as a  comparison with the same
model mentioned above. In this plot,  the data points present a larger
scatter around  the model,  most likely because  the measurement  of a
dispersion is more  affected by the size of the  target sample in each
field. Nonetheless,  the present data  agree well both with  the BRAVA
ones  and with  the model  predictions, for  $|b|$>3.  At  $b=-2$ the
dispersion profile is steeper in the  data than in the model, possibly
because the mass distribution of  the Galactic bulge is more centrally
concentrated than assumed in the model.

%_______________________________________________________________________
\subsection{A kinematical map of the Milky Way}

\begin{figure*}[!ht]
\centering
\includegraphics[width=8.5cm, angle=0]{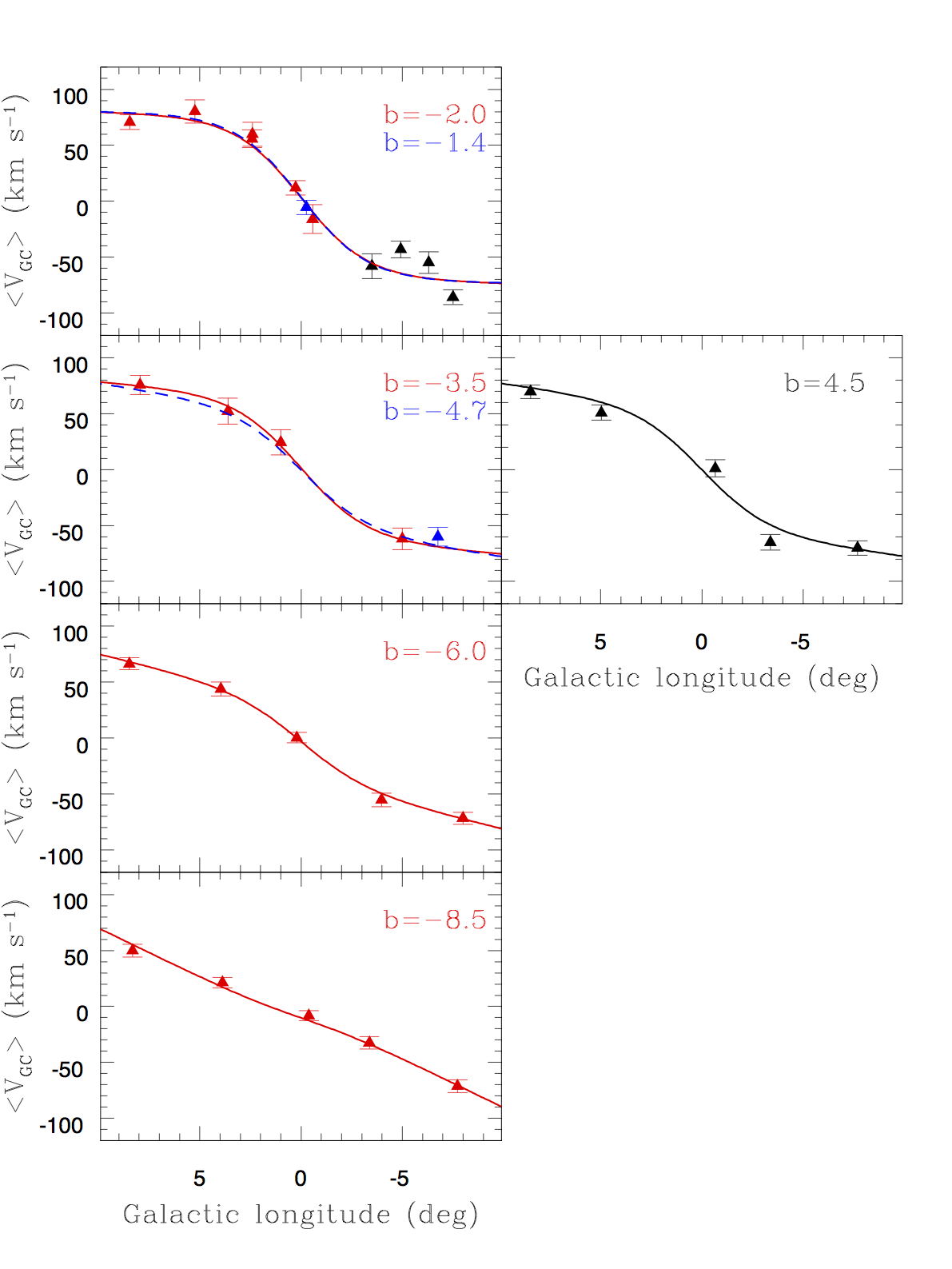}
\includegraphics[width=8.5cm, angle=0]{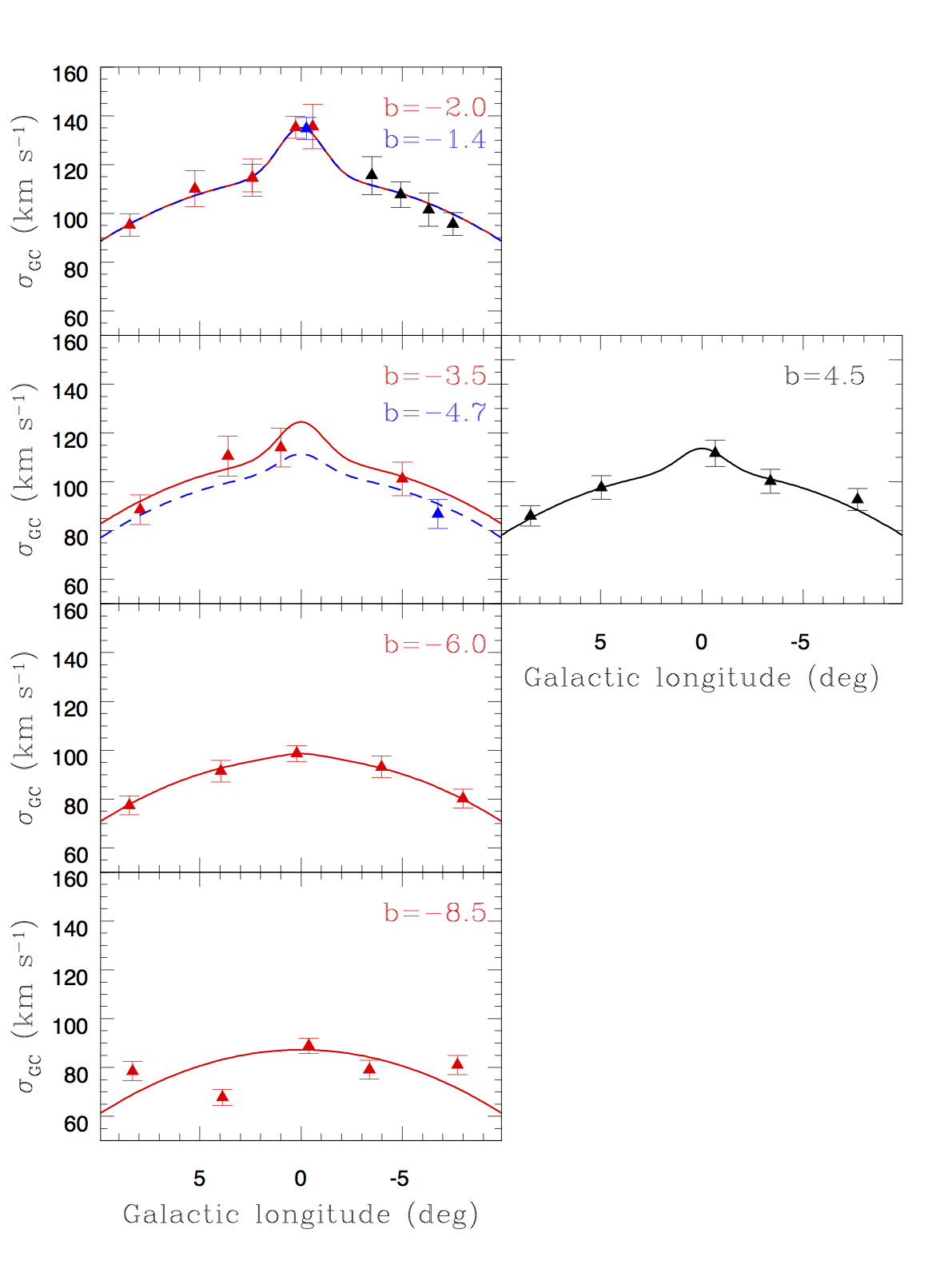}
\caption{Mean  radial  velocity  (two  leftmost  panels)  or  velocity
  dispersion (two rightmost panels) values as a function of longitude,
  at  a  fixed  latitude,  compared  to a  surface  cut  at  the  same
  latitude. Data for a given latitude are shown with the same color of
  the curve  at that latitude.   Black dots  were not included  in the
  fits, for the reasons explained in Sec.~4.2. Points at  $b=+4.5$ are shown here together with  the curve at
  $|b|=4.5$ in order to verify the assumption of symmetry about the
Galactic plane.}
\label{plane_curves}
\end{figure*}

In the following section, we describe the interpolation of radial velocity 
and velocity dispersion at different observed locations, in order to construct 
kinematical maps of
the  Milky Way  bulge.  The  goal of  this exercise  is to  derive the
general rotation  pattern of  the Galactic bulge, to be directly compared
with  kinematic maps  of external  galaxies from  IFU surveys  such as
SAURON,  ATLAS3D   and  CALIFA   \citep{emsellem+2004,  emsellem+2011,
  krajnovic+2011,  husemann+2013}.   In  addition,  it   provides  the
expected $<V_{r}>$ and  $\sigma{V_r}$ values at any location within our
field of view.

\begin{table}
\centering
\caption{Fitted coefficients for equations \ref{RV_eq} and  \ref{SIG_eq}} 
\label{fitting_coeff}
{\small
\begin{tabular}{l l l l}
\hline\hline
\multicolumn{2}{c}{$<V_{GC}>$} & \multicolumn{2}{c}{$\sigma_{RV_{GC}}$} \\
Coeff. & Value & Coeff. & Value \\
\hline
A & $3.80  \pm 1.44$ & A & $79.39  \pm 10.42$ \\
B & $-0.19 \pm 0.03$ & B & $38.45  \pm 9.55$ \\
C & $0.12  \pm 0.01$ & C & $45.51  \pm 19.75$ \\
D & $76.70 \pm 2.58$ & D & $-0.26  \pm 0.01$ \\
E & $-1.17 \pm 0.07$ & E & $21.08  \pm 1.50$ \\
F & $0.30  \pm 0.03$ & r & $523.60 \pm 113.10$ \\
  &                  & s & $2.47   \pm 0.79$ \\
\\
\hline
\end{tabular}
}
\end{table}

In  order  to derive  the  analytical  function, $F(l,b)$,  that  best
describes the  bulge kinematics as a  function of the position  in the
sky, only  data for the  fields at  negative latitude were  used (i.e.
excluding the fields at $b=+4.5$, which are used to test symmetry with
respect to  $b$). The 4  fields at  $l<-3$, $b=-2$ were  also excluded
from the  fit, because they  would introduce an artificial  {\it bump}
that, as explained in Sec. 4.1, is most likely due only the projection
of the bar density along the line of sight.

As a  first step, a function  was fitted to the  radial velocities and
dispersions  at  fixed latitude,  as  a  function of  longitude.   The
functional form {\it is} the same  at every latitude, only the coefficients
{\it were}    allowed   to    change.    A    non   linear,    least-square
Marquardt-Levenberg  fitting algorithm  was used  at this  stage.  The
derived  coefficients were  then plotted  against latitude  and fitted
under the  assumption of symmetry  in $b$.  Combining both  results we
derived  the  set  of  functions   that  can  reproduce  the  observed
kinematics of the inner bulge:

\begin{footnotesize}
\begin{equation}
RV_{GC} = \left( A + B b^{2}\right)  + \left( C b^{2}\right) l + \left( D + E b^{2}\right) \tanh{\left(F l\right)} 
\end{equation} 
\label{RV_eq}
\end{footnotesize}

\begin{footnotesize}
\begin{equation}
\sigma RV_{GC} = \left(A + B e^{-b^{2}/C}\right) + D l^{2} + \left(E e^{-b^{4}/r}\right) e^{-l^{2}/s}.
\end{equation}
\label{SIG_eq} 
\end{footnotesize}
 
Finally,  these functions  were  fitted  again to  all  the fields  at
negative latitude, all  in one step, yielding  the coefficients listed
in  Table~\ref{fitting_coeff}.  Note  that,  when  fitting the  radial
velocity  profile, the  GIBS and  BRAVA data  were combined  given the
excellent   agreement  demonstrated   in  Fig.~\ref{model_rv}.    
Conversely,  when deriving  the velocity  dispersion profile,  only
GIBS data were used because the BRAVA profile is significantly noisier
(Fig.~\ref{model_sig}).

The fits  at different  latitudes are shown  in Fig~\ref{plane_curves}
together with the data used for  the fit ($b<0$; left panels).  In the
right  panels  we  compare  the   fitted  surface  with  the  data  at
$b$=4.5   in  order   to   verify  the  assumption   of
  symmetry. This assumption turns out to be a very accurate both
in radial velocity and dispersion.

\begin{figure}[h]
\centering
\includegraphics[height=8.5cm, angle=0]{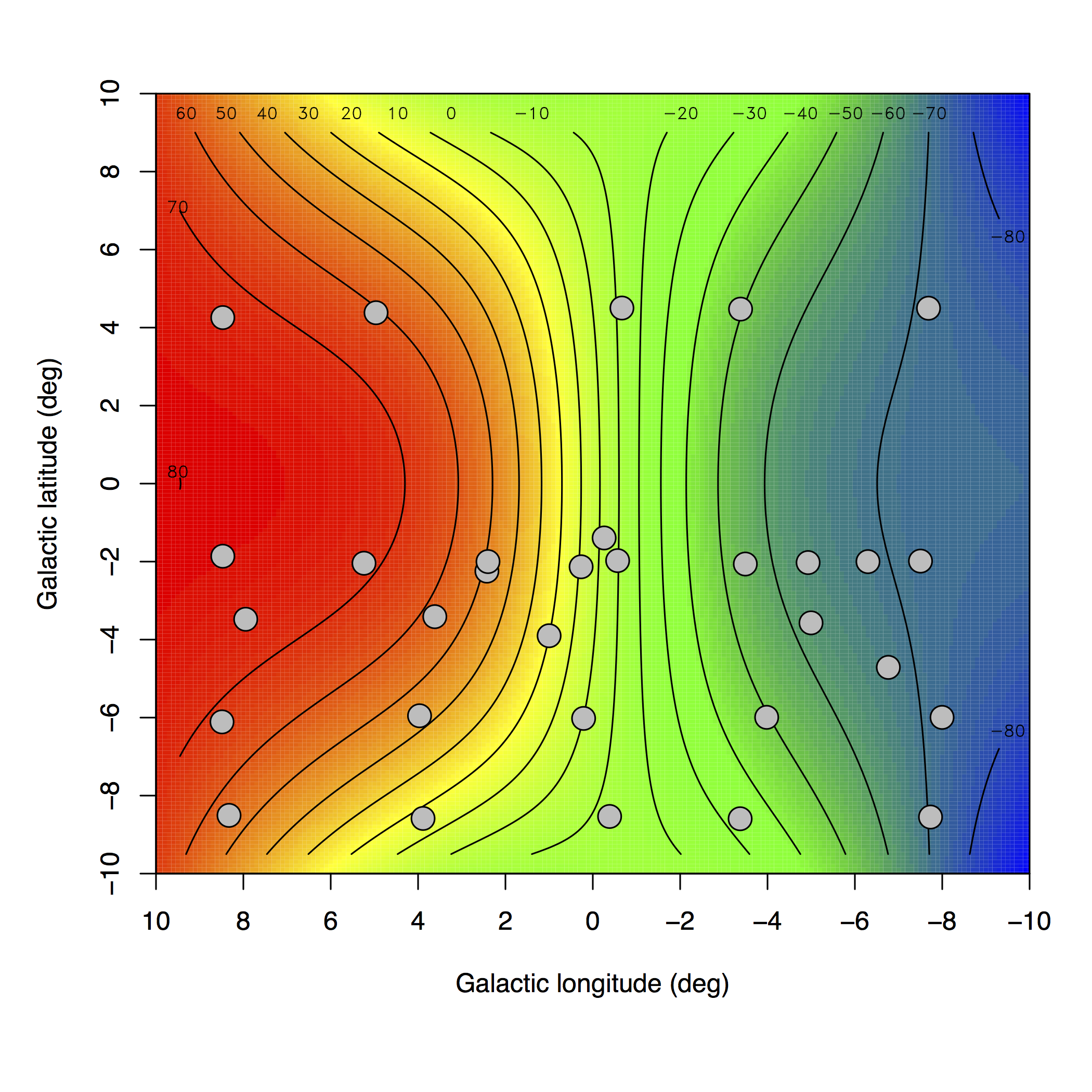}
\caption{Mean radial velocity surface  in the longitude-latitude plane
  constructed  from   the  measured  rotation  profiles   at  negative
  latitudes. Gray points show the positions of the observed fields, while
the black contour lines are labelled with the relevant velocity in km/s.}
\label{plane_rv}
\end{figure}

\begin{figure}[h]
\centering
\includegraphics[height=8.5cm, angle=0]{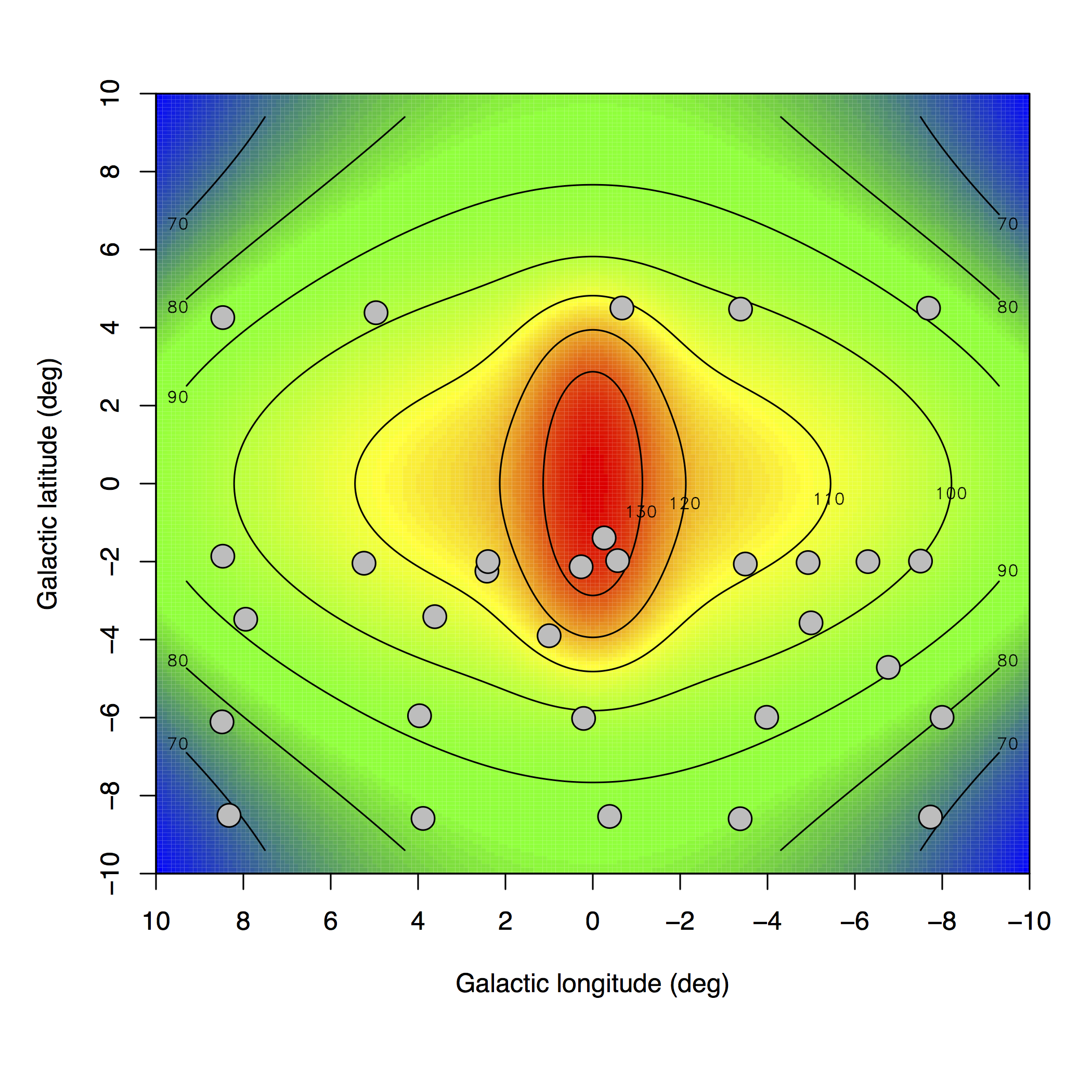}
\caption{Radial velocity dispersion  surface in the longitude-latitude
  plane constructed  from the  measured rotation profiles  at negative
  latitudes. Gray points show the positions of the observed fields, while
the black contour lines are labelled with the relevant velocity 
dispersion in km/s.}
\label{sigma_rv}
\end{figure}

The resulting analytical surfaces are shown in Fig.~\ref{plane_rv} and
Fig.~\ref{sigma_rv}. Two  features of  these maps are  worth noticing.
The first one, in Fig.~\ref{plane_rv},  is the asymmetry of the radial
velocity profile  versus longitude,  which is  only due  to projection
effects,  as demonstrated  by  the agreement  with theoretical  models
shown in Fig.~\ref{model_rv}.  Indeed, the lines of sight crossing the
bar  in its  near side  (positive longitudes)  and far  side (negative
longitudes)  sample  the  bar  density  distribution  in  a  different
way. The second interesting feature is the presence of a {\it peak} in
the   velocity   dispersion,   for    $|l|<2$   and   $|b|<3$   (see
Fig.~\ref{sigma_rv}).  This is  best viewed in Fig.~\ref{plane_curves}
and  it is  due to  the very  large velocity  dispersion in  the three
fields at  ($0,-2$), ($-1,-2$)  and ($0,-1$), as seen in the
histograms in  Fig.~\ref{rv}.  Note  that because in  each of  the two
fields at  ($0,-2$) and  ($0,-1$) we sampled  $\sim$450 stars,  we can
safely exclude the possibility that the  higher velocity dispersion  could be  due to
small  number statistics.  The  peak  is clearly  seen  at $b=-1$  and
$b=-2$,   but it is  not   very   evident  in   the   data   at   $b=-3.5$,
(Fig.~\ref{model_sig}),  possibly due  to the  sparse sampling  of our
fields.   Therefore,  the  elongation  of the  peak  in  the  vertical
direction in  Fig.~\ref{sigma_rv} might not  be real.  While  the real
extension of  the $\sigma$-peak  is not well  constrained, due  to the
presence of only two fields in  that region, we can certainly conclude
that such  a peak  does exist,  and it is  most likely  due to  a high
density peak  in the bulge  innermost region.  It is  interesting that
the spatial extension of the  $\sigma$-peak broadly coincides with the
change in the inclination angle of  the bar, interpreted as due to the
presence of either an inner bar \citep{gonzalez+2011_bar}, or an inner
axisymmetric mass distribution \citep{gerhard+2012}.

%___________________________________________________________________
\subsection{Discussion and Conclusions} 

We  have  presented  the  GIBS Survey,  aimed  at  characterizing  the
kinematics, metallicity  distribution, and  element ratio of  RC stars
across 31 fields in the Galactic  bulge. In addition to describing the
survey in terms  of target selection, observational  strategy and data
products, we provided  here radial velocity measurements  for a sample
of 6392 individual  stars.  The measured velocities agree  well with the
results  from  the   BRAVA  survey,  and  confirm   their  finding  of
cylindrical rotation for  the bulge, extending it  to latitudes $b=-2$
much closer to  the Galactic plane than probed before.  Maps of radial
velocity  and   velocity  dispersions   as  a  function   of  Galactic
coordinates  have been  produced by  interpolating among  the observed
fields. The radial velocity dispersion map is particularly interesting
because it  shows a  central, high  $\sigma$-peak  possibly associated
with  a higher  mass  density  in the  inner  $\sim$2  degrees of  the
Bulge. 

It  is worth  emphasizing  that  the $\sigma$-peak  extends  out to  a
projected distance  of $\sim$280 pc  (corresponding to 2 degrees  at a
distance  of  8  kpc)  and  therefore  it  is   much  bigger  than
(obviously)  the nuclear  star  cluster  surrounding the  supermassive
black hole, but also the other two massive clusters Arches (26 pc from
the Galactic center),  and Quintuplet (30 pc). It is  also much bigger
than  the   Central  Molecular   Zone,  confined  to   $|b|<0.2$  and
$|l|<0.8$.  It does match in size with the change in the position
angle of the Galactic bar \citep{gonzalez+2011_bar} interpreted as
evidence of either a distinct inner bar or of a central axisymmetric
structure \citep{gerhard+2012}. 

Concerning the  bulge formation  scenarios, early merging  of gas-rich
galaxies  and  secular  instabilities   in  stellar  disks  have  been
traditionally regarded as the two  possible channels for the formation
of galactic  bulges \citep[see, e.g.,][]{kormendy+2004}.   Bulges that
would  have formed  by mergers  (often referred  to as  {\it classical
  bulges})  would be  similar to  early  type galaxies  (ETG) in  many
respects,  such  as  old   ages,  $\alpha$-element  enhancement,  high
S\'ersic  index, etc.   Bulges  that  would have  formed  by (bar  and
bar-buckling) instabilities  in a stellar  disk (often referred  to as
{\it pseudobulges}) would be characterized by peanut-shaped isophotes,
lower S\'ersic  index, cylindrical rotation, and  possibly an extended
range  of   stellar  ages   and  lower   $\alpha$-element  enhancement
\citep[e.g.,][and references therein]{shen+2010}.

Our Galactic  bulge fails  to fit  in either  of these  scenarios, but
appears to have some properties  of both scenarios. For example,
it  is  bar-  and  peanut-shaped  and, as  confirmed  by  the  present
investigation, it  rotates cylindrically. Nonetheless, as  reviewed in
the Sec.~1, its  stellar populations appear to be  uniformly $\sim 10$
Gyr  old  \citep{ortolani95,   zoccali+2003,  clarkson+2008}  and  are
$\alpha$     element     enhanced     \citep[e.g.,][and     references
  therein]{gonzalez+2011_alphas, bensby+2013}.  This embarrassment has
prompted the notion  that our bulge may  be a mixture of  both kind of
bulges,  with attempts  at identifying  specific sub-component  of the
bulge with one or the other kind \citep[e.g.,][as already mentioned in
  Sec.1]{hill+2011, babusiaux+2010}.  However,  other groups interpret
multimodal metallicity/kinematical distributions of bulge stars 
uniquely in terms of the disk instability scenario, that may have 
redistributed stars from different disk populations to different 
bulge latitudes, without the need to resort to a contribution of 
merging \citep[e.g.,][]{bensby+2011, bensby+2013, ness+2013b}.

There are, however, lines of evidence suggesting that Nature may
have  followed  also  other  paths   for  the  formation  of  galactic
bulges. First,  the notion according  to which ETGs are  primarily the
result of  merging is seriously  challenged by the finding  that $\sim
86\%$  of them  are fast  rotators, often  with cylindrical  rotation,
whereas only $\sim 14\%$ of them  are slow rotators, the likely result
of  dry merging \citep{emsellem+2011}.  Independent evidence  for a
marginal role  of merging in  shaping passively evolving  (early type)
galaxies  comes from  the characteristic  mass ($M_*$)  of their  mass
function  being only  $\sim 0.1$  dex higher  in high-density  regions
(where merging takes place) compared  to low-density regions \citep{peng+2012}.

The second evidence comes from  high-redshift galaxies. Indeed, if the
bulk of  star in  the bulge are  $\sim 10$  Gyr old, then  it is  at a
$\sim$10 Gyr lookback  time that we should look to  see analogs of our
bulge in formation, i.e., at $z\sim  2$.  A great deal of evidence has
accumulated  in recent  years on  star-forming galaxies  at such  high
redshifts.   Many  among  them   are  large,  clumpy,  rotating  disks
\citep{genzel+2006,  forster+2009}  with  high  star  formation  rates
\citep[SFR, e.g.,][]{daddi+2007}.  Moreover, such  disks are much more
gas rich  compared to local  spirals, with  gas fractions of  order of
$\sim 50\%$ or more \citep{tacconi+2010, daddi+2010}. These properties
makes them very attractive in the  context of bulge formation.  On the
one hand, their high specific SFR ($\equiv$SFR/$M_{\rm star}$), larger
than  the   corrsponding  inverse  Hubble  time   (i.e.,  $>  10^{-9}$
Gyr$^{-1}$) implies a  very rapid mass growth  and automatically leads
to $\alpha$-element enhanced  stellar populations \citep{renzini+2009,
  peng+2010,  lilly+2013}.  Furthermore,  such  gas-rich galaxies  are
prone to disk instabilities leading  to massive clump formation, which
in turn can migrate to the center and dissipatively coalesce resulting
in   bulge   formation   over   timescales  of   a   few   $10^8$   yr
\citep[e.g.,][]{immeli+2004,        carollo+2007,       elmegreen+2008,
  bournaud+2009}.   Such  timescales  are   much  shorter  than  those
typically ascribed to  secular instabilities in local  (gas poor) disk
galaxies. It should  go without saying that bulges formed  in this way
would  be relatively  fast rotators  and  may later  develop bars  and
X-shaped components. It is worth emphasizing that this may well be the
dominant channel  for the formation  of both galactic bulges  and most
ETGs  alike, with  what remains  to be  understood being  the physical
processes  leading to  the quenching  of their  star formation.   Also
worth emphasizing, is the fundamental  difference between this kind of
disk instability, which is intimately  related to disks being very gas
rich,  and the  traditional  {\it  secular}, bar/buckling  instability
\citep[\`a  la][]{sellwood+1981} which  instead develops  in a  purely
stellar disk.

For this  reason, we  refrain from  intepreting the  bulge cylindrical
rotation as an  evidence of its RC stars tracing  a {\it pseudo}-bulge
component, or pointing towards its formation via {\it secular} 
evolution.

\smallskip

%___________________________________________________________________
\begin{acknowledgements}
  SV and MZ acknowledge support  by Proyecto Fondecyt Regular 1110393,
  by  the BASAL  Center for  Astrophysics and  Associated Technologies
  PFB-06, the FONDAP Center for Astrophysics 15010003, Proyecto Anillo
  ACT-86 and by the Chilean Ministry for the Economy, Development, and
  Tourism's Programa  Iniciativa Cient\'{i}fica Milenio  through grant
  P07-021-F, awarded to The Milky Way Millennium Nucleus.
\end{acknowledgements}

%___________________________________________________________________
\bibliographystyle{aa}
\bibliography{mybiblio}

\end{document}